\documentclass[conference]{IEEEtran}
\IEEEoverridecommandlockouts

\usepackage{cite}
\usepackage{amsmath,amssymb,amsfonts}
\usepackage{graphicx}
\usepackage{textcomp}
\usepackage{xcolor}
\usepackage{booktabs}
\usepackage{tabularx}
\usepackage{subcaption}
\usepackage{float}
\usepackage{algorithm}
\usepackage{algpseudocode}
\usepackage{amsthm}
\usepackage{hyperref}
\hypersetup{hidelinks}
\usepackage{pifont}
\usepackage{microtype}
\usepackage{balance}

\newtheorem{proposition}{Proposition}
\newcommand{\centperkwh}{\textcent/kWh}
\begin{document}

\title{\texttt{BOOST:} Microgrid Sizing using Ordinal Optimization
\thanks{979-8-3315-4112-5/25/\$31.00 ~\copyright~2025 IEEE}}

\author{
    \IEEEauthorblockN{Mohamad Chehade\IEEEauthorrefmark{1}, Sami Karaki\IEEEauthorrefmark{2}}
    \IEEEauthorblockA{
        \IEEEauthorrefmark{1}Chandra Department of Electrical and Computer Engineering, The University of Texas at Austin, Austin, TX, USA \\ chehade@utexas.edu
    }
    \IEEEauthorblockA{
        \IEEEauthorrefmark{2}Electrical and Computer Engineering Department, American University of Beirut, Beirut, Lebanon \\ skaraki@aub.edu.lb
    }
}

\maketitle

\begin{abstract}
Sizing a residential microgrid efficiently requires solving a coupled design-and-operation problem: photovoltaic (PV) and battery capacities should be chosen in a way that reflects how the system will actually be dispatched over time. This paper proposes \texttt{BOOST}, or \emph{Battery-solar Ordinal Optimization Sizing Technique}, which combines ordinal optimization (OO) with mixed-integer linear programming (MILP). OO is used to screen a large set of candidate battery/PV designs with a simple linear model and then re-evaluate only the most promising designs with a more accurate MILP that captures diesel-commitment logic. Relative to the original short paper, this expanded manuscript retains the full methodological narrative but refreshes the quantitative section using a new synthetic benchmark dataset suite generated from the released clean reimplementation. The suite contains five yearly synthetic datasets/configurations---base, cheap battery, cheap PV, expensive diesel, and high peak tariff. On the base synthetic dataset, the best accurate design is a 500~kWh battery with 1833.3~kW of PV, achieving 13.169~\textcent/kWh, while \texttt{BOOST} improves upon dynamic programming and greedy baselines. Across the full $10\times10$ design grid, the LP and MILP rankings are effectively identical ($\rho=1.000$), the paper-style choice of $N=90$ and $s=18$ recovers the global accurate optimum, and the OO-based workflow reduces runtime by 51.8\% relative to exhaustive accurate evaluation on the refreshed synthetic benchmark run. Because these added datasets are synthetic, they should be read as methodology-validation evidence. Code is available at \url{https://github.com/MFHChehade/Microgrid-Optimization}.
\end{abstract}


\section{Introduction}
\label{sec:introduction}
Power systems are under increasing pressure to incorporate low-carbon and resilient energy resources. International climate policy and unreliable distribution grids in some regions both motivate local generation and storage deployment \cite{Paris_Agreement,Outage,LebanonOutages}. In this setting, residential and community microgrids built from photovoltaic (PV) generation, batteries, diesel backup, and grid interconnection have become an important design problem \cite{sizing_motivation,Hybrid2}.

Sizing and operation are tightly coupled: a battery/PV configuration is only attractive if it can be dispatched economically over time. Exact MILP-based co-optimization can model this well, but repeated MILP evaluation across a large design space can be expensive \cite{MILP_computationally_expensive,Mehrtash1,Mehrtash2,MILP_wu2021}. Heuristic and metaheuristic methods can scale better, but may require many evaluations and offer limited guarantees \cite{Sawwas,GA_campus2020,PSO,RETScreen}.

Ordinal optimization (OO) offers an appealing middle ground. Rather than insisting on perfectly accurate evaluation for every candidate design, OO exploits the fact that preserving the \emph{order} of promising designs is often enough to identify a good-enough solution efficiently \cite{Ho2000,Ho2007,Jia2006}. This perspective has already been used in power-system applications such as distributed-generation placement, meter placement, and renewable-energy sizing \cite{Jabr,Singh,OO_power,Majed,Dinnawi,Zein}. However, in several OO-based renewable-energy sizing workflows, the inner operational problem is handled by greedy or dynamic-programming (DP) methods, which may discretize the battery state of charge and become memory-intensive as the state space grows \cite{Richa,Zein,DP_battery_2018,DP_microgrid_2022}. These approximations can degrade dispatch quality even when the outer OO logic is well motivated.

This work proposes a Battery-solar Ordinal Optimization Sizing Technique (\texttt{BOOST}) for a microgrid similar to that studied in \cite{Richa}. The microgrid serves a residential load using the utility grid, a diesel generator, a photovoltaic array, and battery storage. Relative to the greedy one-step-ahead strategy in \cite{Richa}, our key modeling change is to solve the inner operational problem with mixed-integer linear programming. The resulting MILP explicitly optimizes charging, discharging, grid purchases, and diesel use over the full operating horizon, yielding a more principled operational cost for each candidate size.

In brief, our main contributions are:
\begin{enumerate}
    \item We propose a PV--battery microgrid sizing technique that combines ordinal optimization (OO) with mixed-integer linear programming (MILP).
    \item We refresh the quantitative study using a new synthetic benchmark dataset suite generated from the released clean reimplementation, including five yearly dataset/configuration variants.
    \item We show on these new datasets that LP screening preserves the accurate ranking of promising designs and that the sampled \texttt{BOOST} optimum matches exhaustive accurate search on the full grid.
    \item We numerically show that the MILP-based inner solver improves upon dynamic-programming and greedy alternatives on the base synthetic dataset, while additional datasets quantify sensitivity to battery, PV, diesel, and tariff assumptions.
\end{enumerate}

The rest of the paper is organized as follows. Section~\ref{sec:preliminaries} introduces the microgrid setting and the questions of interest, Section~\ref{sec:inner_problem} formulates the inner operational problem, Section~\ref{sec:ordinal_optimization} discusses the proposed OO algorithm, Section~\ref{sec:experiments} presents the new synthetic dataset suite and refreshed numerical evidence, and Section~\ref{sec:conclusion} concludes the paper.

\section{Preliminaries}
\label{sec:preliminaries}
\begin{figure*}[!t]
    \centering
    \begin{subfigure}[t]{0.33\textwidth}
        \centering
        \includegraphics[width=1.2\textwidth]{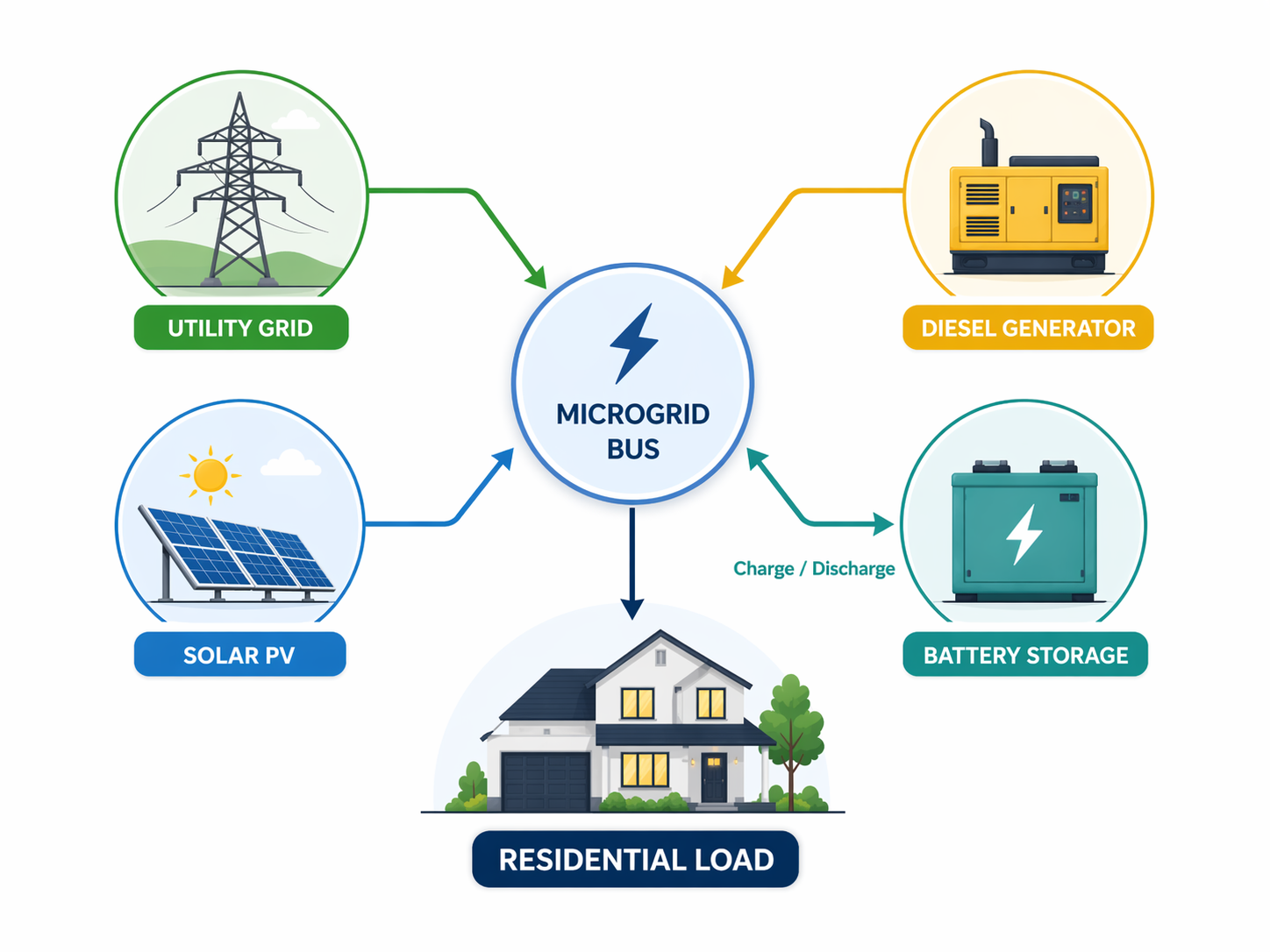}
        \caption{Physical microgrid components.}
        \label{fig:system_components}
    \end{subfigure}\hfill
    \begin{subfigure}[t]{0.60\textwidth}
        \centering
        \includegraphics[width=1\textwidth]{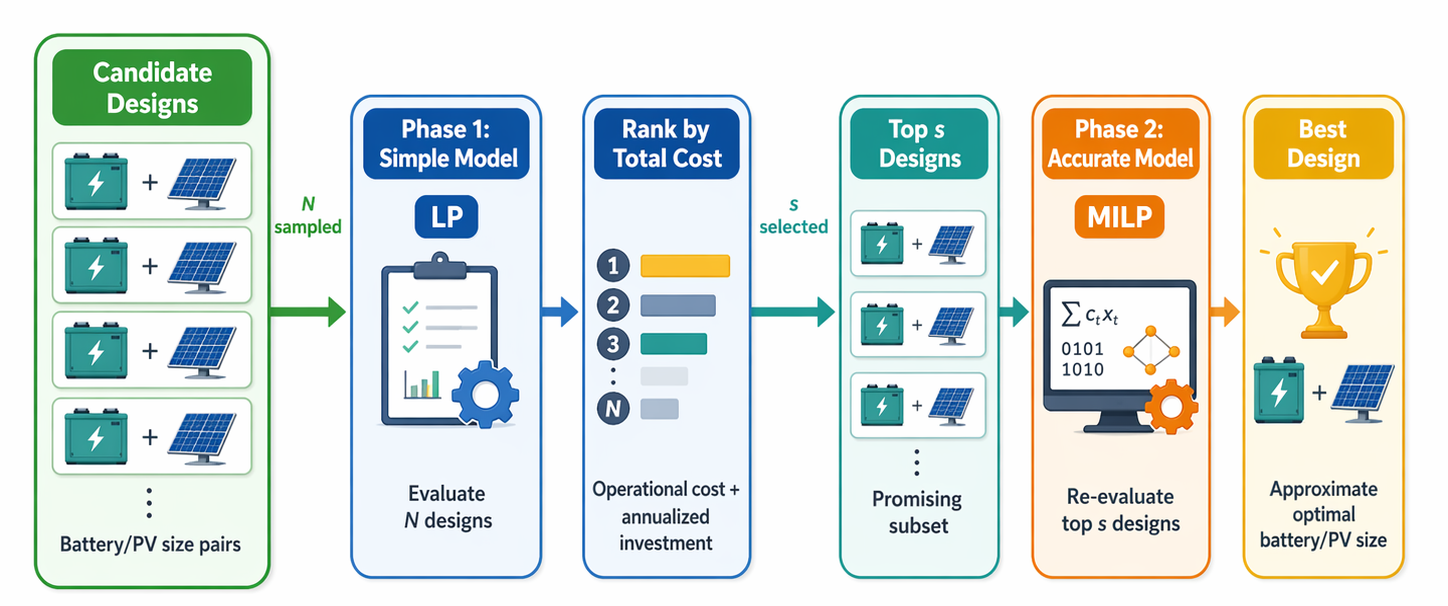}
        \caption{\texttt{BOOST} screening and re-evaluation workflow.}
        \label{fig:OO}
    \end{subfigure}
    \caption{System setting and algorithmic overview. The microgrid contains five interacting components, while \texttt{BOOST} uses a two-phase OO procedure to identify promising battery/PV sizes without evaluating every design with the expensive accurate model.}
    \label{fig:system_and_workflow}
\end{figure*}

We consider the microgrid shown in Fig.~\ref{fig:system_and_workflow}. A load with demand $d_t$ is supplied with electricity from 1) the grid with a time-varying price $p_{G,t}$ (in \$/kWh), 2) a diesel generator with capacity $E_D$ (in kWh) and fixed price $p_D$ (in \$/kWh), 3) a battery with capacity $E_B$ (in kWh), and 4) a solar photovoltaic (PV) panel with maximum power output $P^{\text{max}}_{\text{PV}, t}$ at time $t$. Assuming that the grid and the diesel generator have already been connected, this work aims to answer the following two questions:

\textit{(Q1) What are the optimal sizes for the battery and PV panels that minimize energy consumption from the grid and diesel generator while keeping investment costs low?}

\textit{(Q2) What is the power output of each of the energy resources at a given time instance?}

The left panel of Fig.~\ref{fig:system_and_workflow} shows the physical components of the microgrid. The right panel shows the logic of \texttt{BOOST}: many candidate designs are screened using a simple model, then only the most promising designs are re-evaluated using the accurate model. Keeping both diagrams together clarifies the distinction between the physical system and the OO workflow that acts on top of it.

\section{The Inner Problem}
\label{sec:inner_problem}
To answer (Q2), i.e., to find the contribution of each resource at a given time $t$, we formulate the problem as a linear program (LP). The objective in \eqref{eq:objective_function} is to minimize the cost of electricity purchased from the grid and the diesel generator.

\begin{subequations} \label{eq:LP}
    \begin{align}
        &\text{minimize} \quad \sum_{t} p_{\text{G}}(t) P_{\text{G}}(t) + p_{\text{D}}(t) P_{\text{D}}(t) \tag{1a} \label{eq:objective_function}\\
        &\text{subject to} \notag \\
        &\quad P_{\text{PV}}(t) + P_{\text{D}}(t) + P_{\text{G}}(t) + P_{\text{B,disch}}(t) = P_{\text{L}}(t) + P_{\text{B,ch}}(t) \tag{1b} \label{eq:power_balance}\\
        &\quad 0 \leq P_{\text{PV}}(t) \leq P_{\text{PV}}^{\text{max}} \tag{1c} \label{eq:pv_limits}\\
        &\quad \text{SOC}(t+1) = \text{SOC}(t) + \eta_{\text{ch}} P_{\text{B,ch}}(t) - \frac{1}{\eta_{\text{disch}}} P_{\text{B,disch}}(t) \tag{1d} \label{eq:soc_dynamics}\\
        &\quad \text{SOC}_{\text{min}} \leq \text{SOC}(t) \leq \text{SOC}_{\text{max}} \tag{1e} \label{eq:soc_limits}\\
        &\quad 0 \leq P_{\text{D}}(t) \leq P_{\text{D}}^{\text{max}} \tag{1f} \label{eq:diesel_limits}
    \end{align}
\end{subequations}

Equation~\eqref{eq:power_balance} describes the power balance. The upper and lower limits on PV output, diesel-generator output, and battery state of charge (SOC) are expressed in \eqref{eq:pv_limits}, \eqref{eq:diesel_limits}, and \eqref{eq:soc_limits}, respectively. Finally, \eqref{eq:soc_dynamics} describes the battery dynamics, where $\eta_{\text{ch}}$ and $\eta_{\text{disch}}$ denote charging and discharging efficiencies.

The optimization problem above is referred to as the \emph{simple model}. A more accurate model requires that if the diesel generator is used, its power output must be above a minimum threshold $P_D^{\text{min}}$. Adding this logic introduces a binary variable $u_D(t)$ and turns the problem into a mixed-integer linear program (MILP):

\begin{subequations} \label{eq:MILP}
    \begin{align}
        &\text{minimize} \quad \sum_{t} p_{\text{G}}(t) P_{\text{G}}(t) + p_{\text{D}}(t) P_{\text{D}}(t) \tag{2a} \label{eq:new_minimize}\\
        &\text{subject to} \notag \\
        &\quad \eqref{eq:power_balance}-\eqref{eq:soc_limits} \notag \\
        &\quad P_{\text{D}}^{\text{min}} u_{\text{D}}(t) \leq P_{\text{D}}(t) \leq P_{\text{D}}^{\text{max}} u_{\text{D}}(t) \tag{2b} \label{eq:new_constraint_PD}
    \end{align}
\end{subequations}

We refer to \eqref{eq:MILP} as the \emph{accurate model}. The simple LP is cheaper to solve and is therefore used for large-scale screening, whereas the accurate MILP is reserved for the final re-evaluation of a small subset of designs.

\section{Ordinal Optimization (OO)}
\label{sec:ordinal_optimization}
Problem \eqref{eq:LP} finds the minimal operational cost of the system \emph{given} fixed battery and PV sizes. Since we are interested in the \emph{optimal} sizes of the battery and PV systems, we could in principle solve the accurate MILP in \eqref{eq:MILP} for every possible combination $(E_{\text{B}}, \text{PV}_{\text{size}})$. However, given the large number of such combinations and the fact that MILPs are NP-hard, exhaustive accurate evaluation can be expensive. To accelerate the process, we instead screen designs using the approximate LP in \eqref{eq:LP}. The central premise is summarized by the following proposition, consistent with the classical OO viewpoint that order information can be substantially cheaper to obtain than exact value information \cite{Ho2000,Ho2007,Jia2006}.

\begin{proposition} \label{prop:robust_order}
Order is robust with respect to estimation noise.
\end{proposition}

In other words, if a set of designs is evaluated using the accurate MILP model and ordered in terms of cost, their order should not differ much when they are evaluated using the simple LP model. The costs are less precise under the LP, but the overall \emph{rank} should not be drastically affected.

Our OO approach to the sizing problem is divided into two phases: (1) evaluate a large set of possible designs using the simple LP model and rank them in terms of overall cost; (2) extract a subset of the top designs found using the LP model and re-evaluate them using the accurate MILP model. Crucial to this approach is determining $N$, the number of designs to be evaluated using the simple model, and $s$, the number of top designs to be re-evaluated using the accurate model.

\subsection{Phase 1: The Simple Model}
We want to sample enough designs so that at least one of the sampled designs is a top design. Therefore, we choose $N$ such that at least one of the $N$ designs lies in the actual top-$\alpha$\% designs with probability $P$:
\begin{equation}
\label{eq:N}
    N \geq \frac {\ln(1-P)} {\ln(1-\alpha)}
\end{equation}

We solve the simple LP model for each of the $N$ designs, thereby finding the associated operational cost. The overall cost of a design is then the sum of the operational cost and the annualized investment costs of the solar and battery systems:
\begin{align}
\text{Total Cost} &= \underbrace{\text{Op. Cost}}_{\text{using LP}} + \underbrace{\text{Inv. Cost}_{\text{PV}} + \text{Inv. Cost}_{\text{Battery}}}_{\text{annualized via annuity}} \label{eq:total_cost}
\end{align}
The $N$ designs are ranked based on this total cost, from which the top $s < N$ designs are chosen for accurate re-evaluation.

\subsection{Phase 2: The Accurate Model}
While \eqref{eq:N} guarantees with probability $P$ that a good-enough design $(E_B,\,\text{PV}_{\text{size}})^*$ exists among the $N$ designs, Proposition~\ref{prop:robust_order} claims that $(E_B,\,\text{PV}_{\text{size}})^*$ should also be highly ranked by the simple model. Therefore, we only need to search through a set of top designs of size $s$.

If $S$ denotes the top-$s$ designs obtained from the simple model and $G$ denotes the set of truly good solutions of size $g$, then the probability that $S$ and $G$ have at least $k$ common elements is
\begin{equation}
\label{eq:s}
AP(k) = \Pr(|G \cap S| \geq k) = \sum_{i=k}^{\min(g,s)}
\frac{\binom{g}{i} \binom{N-g}{s-i}} {\binom{N}{s}}.
\end{equation}

$AP(k)$ is the \emph{alignment probability}. We choose the minimum possible $s$ that achieves a desired alignment probability, then solve the accurate MILP for each of the top-$s$ designs and recompute the total cost as in \eqref{eq:total_cost}. The design with minimum cost is taken to be the best approximation of $(E_B,\,\text{PV}_{\text{size}})^*$.

Our approach, denoted by \texttt{BOOST}, is summarized in Algorithm~\ref{alg:BOOST}.

\begin{algorithm}[!t]
\caption{\texttt{BOOST}}
\label{alg:BOOST}
\begin{algorithmic}[1]
\State Compute $N$ using \eqref{eq:N}
\For{$i \gets 1$ to $N$}
    \State Solve the simple LP problem in \eqref{eq:LP}
    \State Compute total cost using \eqref{eq:total_cost}
\EndFor
\State Rank the $N$ designs by total cost
\State Compute $s$ using \eqref{eq:s}
\State Select the top $s$ designs from the $N$ candidates
\For{$j \gets 1$ to $s$}
    \State Solve the accurate MILP problem in \eqref{eq:MILP}
    \State Compute total cost using \eqref{eq:total_cost}
\EndFor
\State Rank the $s$ designs by total cost
\State \Return $\left(E_B,\text{PV}_{\text{size}}\right)_1 \approx \left(E_B,\text{PV}_{\text{size}}\right)^*$
\end{algorithmic}
\end{algorithm}

\section{Numerical Experiments}
\label{sec:experiments}
\subsection{Synthetic Benchmark Dataset Suite and Experimental Protocol}
The expanded manuscript reports results on a five-case synthetic benchmark suite generated from the released clean reimplementation: \emph{base}, \emph{cheap battery}, \emph{cheap PV}, \emph{expensive diesel}, and \emph{high peak tariff}. Each case uses one year of hourly traces, the same microgrid structure, and weekly rolling-horizon evaluation. The candidate design space is a $10\times10$ battery/PV grid, and the OO parameters use the paper-style setting $N=90$ and $s=18$.

The new experiments evaluate screening fidelity, runtime savings versus exhaustive accurate evaluation, and sensitivity of the selected design and LCOE across the dataset suite. In addition, the benchmark is structured so that each synthetic case preserves the same optimization pipeline while changing only the economic context, which isolates ranking robustness from any one cost profile or dispatch week.

From a methodological standpoint, this suite is not intended to replace the original application motivation with synthetic evidence. Instead, it creates controlled annual cases that isolate how the OO screen behaves when asset costs and tariff structure change. This makes it possible to test not only whether \texttt{BOOST} finds one good design, but also whether the LP-based ranking remains stable when the economic ordering of candidate designs shifts across scenarios.

Table~\ref{tab:new_primary_results} summarizes the base synthetic dataset. The best accurate design is a 500~kWh battery paired with 1833.3~kW of PV, with accurate LCOE 13.169~\centperkwh. The LP and accurate rankings are essentially identical (Spearman $\rho=1.000$), $s=18$ recovers the full top-10 accurate set and contains the global accurate optimum, and the OO workflow reduces runtime from 169.9~s to 82.0~s (51.8\%).

\begin{table}[!t]
\caption{Base synthetic dataset: primary quantitative results from the refreshed synthetic benchmark.}
\label{tab:new_primary_results}
\centering
\small
\begin{tabularx}{\columnwidth}{@{}>{\raggedright\arraybackslash}p{0.64\columnwidth}>{\raggedleft\arraybackslash}p{0.28\columnwidth}@{}}
\toprule
\multicolumn{2}{@{}l}{\textbf{Best design and cost}} \\
Best accurate design $(E_B,\mathrm{PV})$ & $(500~\text{kWh},\;1833.3~\text{kW})$ \\
Best accurate LCOE & 13.169~\centperkwh \\
\addlinespace[2pt]
\multicolumn{2}{@{}l}{\textbf{Screening fidelity}} \\
Spearman rank correlation $\rho$ & 1.000 \\
Recall of true top-10 at $s=18$ & 100\% \\
Sampled \texttt{BOOST} recovers global optimum & Yes \\
\addlinespace[2pt]
\multicolumn{2}{@{}l}{\textbf{Runtime}} \\
LP screen, 100 designs & 53.2~s \\
Exhaustive accurate, 100 designs & 169.9~s \\
\texttt{BOOST}, 90 LP + 18 MILP & 82.0~s \\
Reduction vs. exhaustive accurate & 51.8\% \\
\bottomrule
\end{tabularx}
\end{table}

Table~\ref{tab:new_baselines} compares operational approaches on the same base synthetic dataset. The accurate inner solver gives the best LCOE, the gap to DP is small but consistent, and the greedy policy is noticeably worse.

\begin{table}[!t]
\caption{Operational comparison on the base synthetic dataset.}
\label{tab:new_baselines}
\centering
\begin{tabular}{lcc}
\toprule
Method & LCOE (\centperkwh) & Gap vs. \texttt{BOOST} \\
\midrule
\textbf{\texttt{BOOST}} & \textbf{13.169} & -- \\
DP & 13.178 & +0.009 \\
Greedy & 13.251 & +0.082 \\
\bottomrule
\end{tabular}
\end{table}

Table~\ref{tab:scenario_results} reports the best design selected on each synthetic dataset. Cheaper battery economics push the battery to 5000~kWh, cheaper PV pushes the PV size to 2500~kW, while the expensive-diesel and high-peak-tariff datasets keep the base design but change the achieved LCOE.

\begin{table*}[!t]
\caption{Best design and LCOE across the new synthetic dataset suite.}
\label{tab:scenario_results}
\centering
\begin{tabular}{lccc}
\toprule
Synthetic dataset & Best battery size (kWh) & Best PV size (kW) & Accurate LCOE (\centperkwh) \\
\midrule
Base & 500 & 1833.3 & 13.169 \\
Cheap battery & 5000 & 2500.0 & 12.985 \\
Cheap PV & 500 & 2500.0 & 12.190 \\
Expensive diesel & 500 & 1833.3 & 14.191 \\
High peak tariff & 500 & 1833.3 & 13.169 \\
\bottomrule
\end{tabular}
\end{table*}

\begin{figure*}[!t]
    \centering
    \begin{subfigure}[t]{0.31\textwidth}
        \centering
        \includegraphics[width=\textwidth]{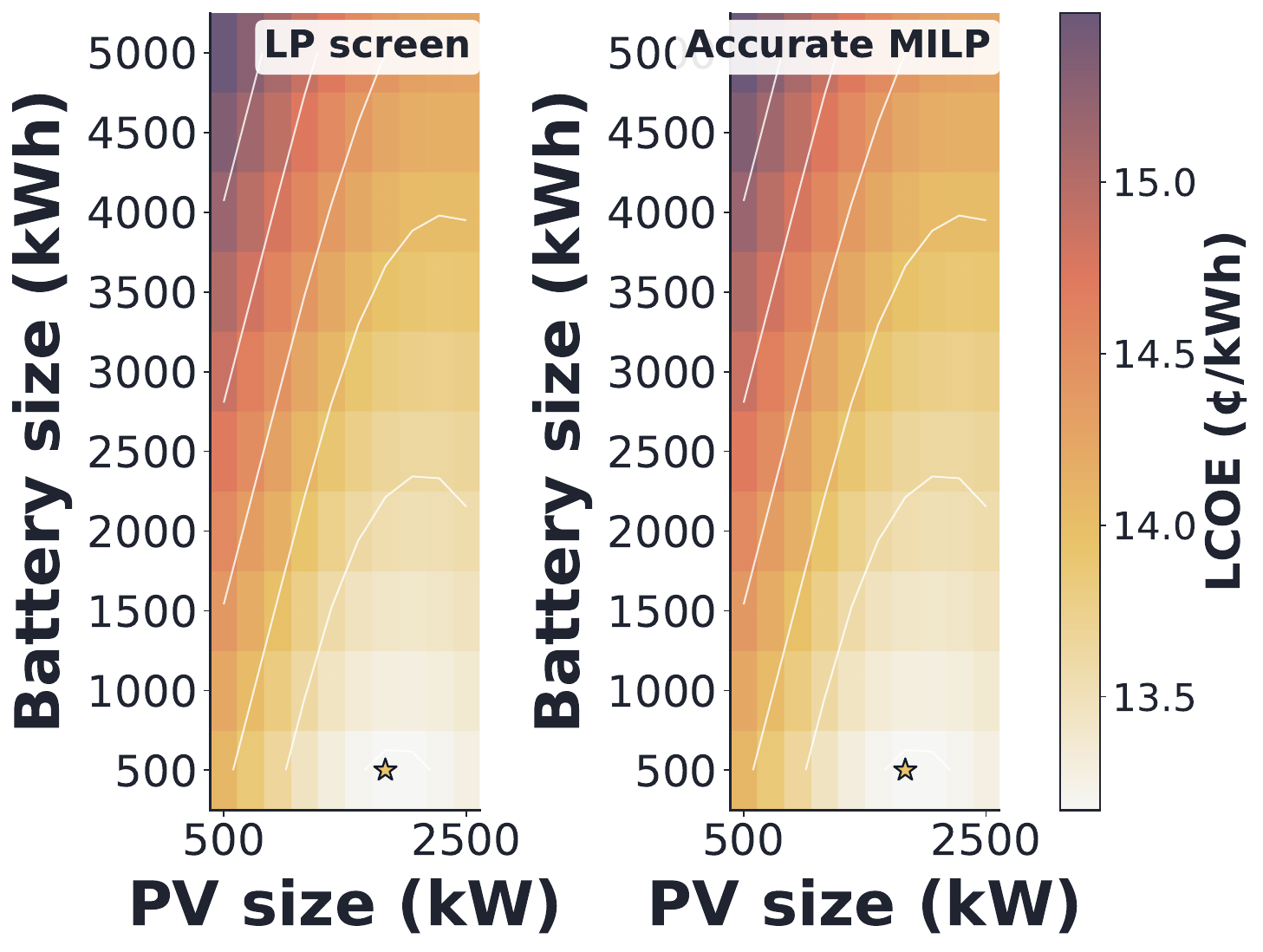}
        \caption{LP and accurate design landscapes.}
        \label{fig:landscape}
    \end{subfigure}\hfill
    \begin{subfigure}[t]{0.31\textwidth}
        \centering
        \includegraphics[width=\textwidth]{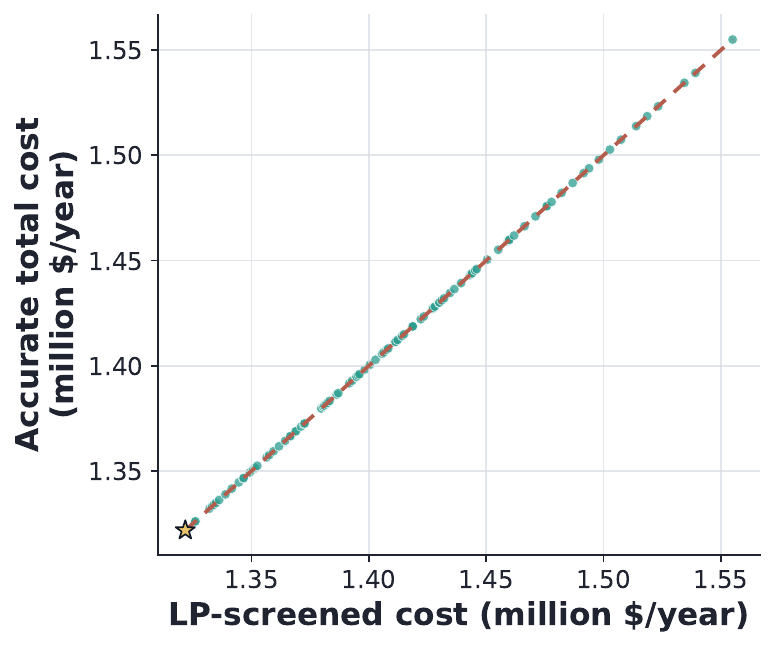}
        \caption{LP versus accurate total cost.}
        \label{fig:scatter}
    \end{subfigure}\hfill
    \begin{subfigure}[t]{0.31\textwidth}
        \centering
        \includegraphics[width=\textwidth]{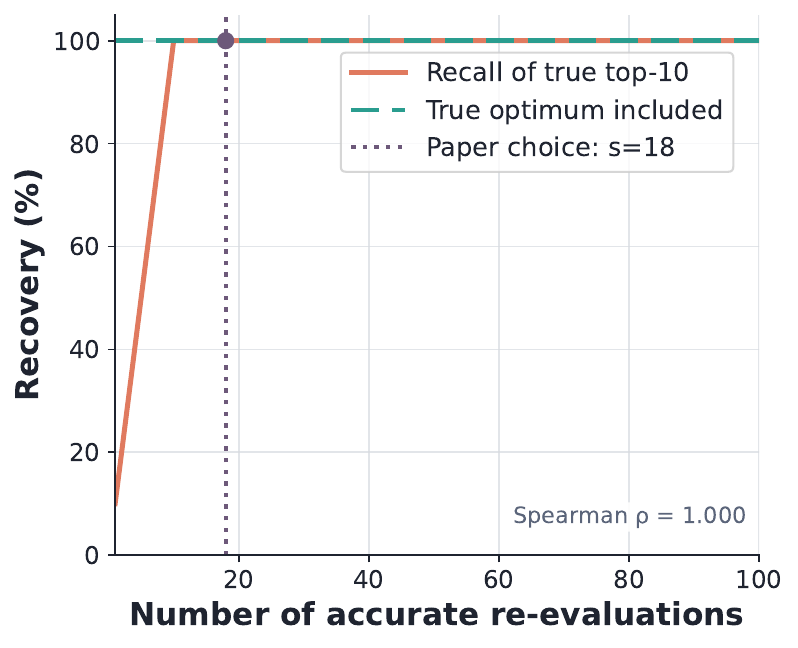}
        \caption{Recall as accurate re-evaluations increase.}
        \label{fig:screening}
    \end{subfigure}
    \caption{Expanded synthetic benchmark, row 1: screening fidelity across the new synthetic datasets. The LP and accurate landscapes remain aligned across the full design grid, LP and accurate total costs stay close over all candidate designs, and the retained set reaches full top-10 recall at $s=18$.}
    \label{fig:row1}
\end{figure*}

\begin{figure*}[!t]
    \centering
    \begin{subfigure}[t]{0.31\textwidth}
        \centering
        \includegraphics[width=\textwidth]{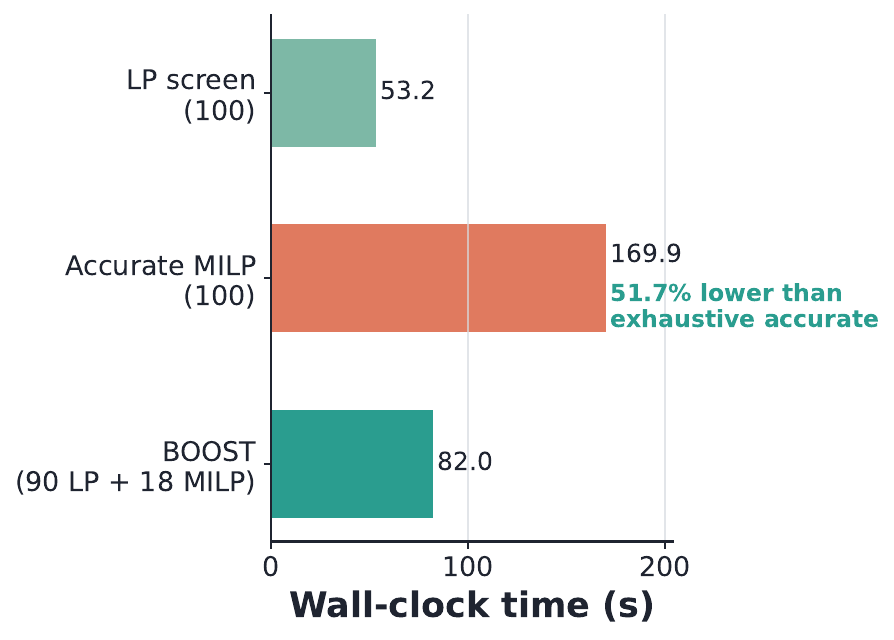}
        \caption{Runtime for exhaustive and OO-based evaluation.}
        \label{fig:runtime}
    \end{subfigure}\hfill
    \begin{subfigure}[t]{0.31\textwidth}
        \centering
        \includegraphics[width=\textwidth]{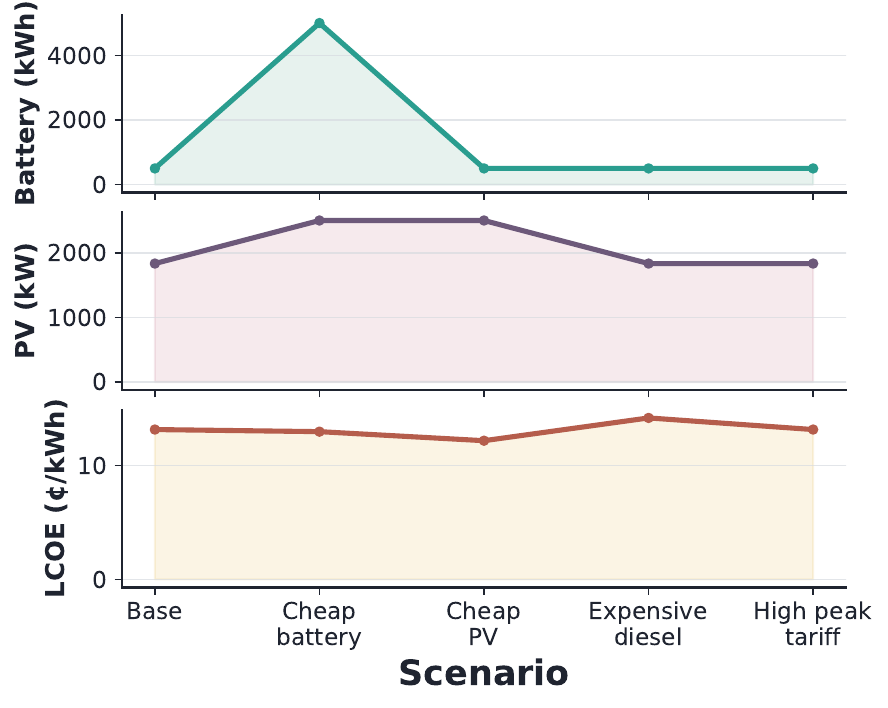}
        \caption{Selected sizes and LCOE across scenarios.}
        \label{fig:scenario}
    \end{subfigure}\hfill
    \begin{subfigure}[t]{0.31\textwidth}
        \centering
        \includegraphics[width=\textwidth]{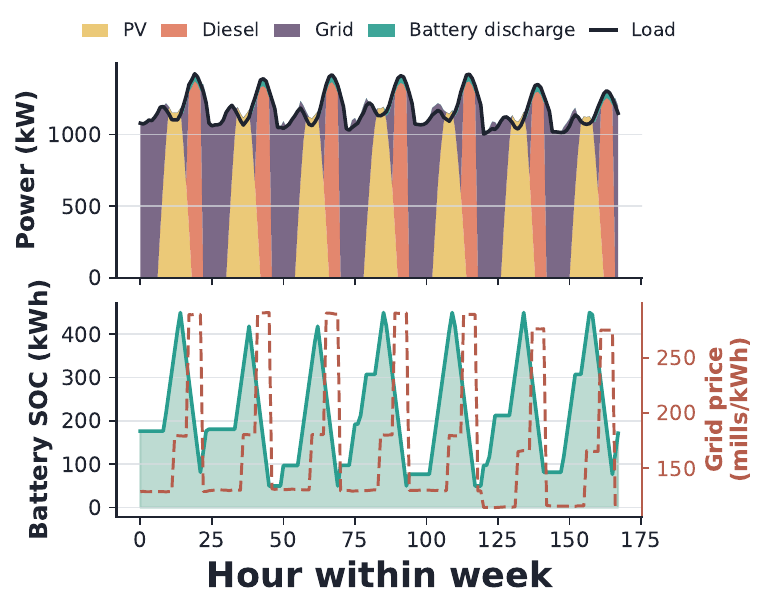}
        \caption{Representative summer-week dispatch.}
        \label{fig:dispatch}
    \end{subfigure}
    \caption{Expanded synthetic benchmark, row 2: runtime, sensitivity, and operational interpretation. The OO-based workflow reduces accurate-evaluation time by 51.8\% on the refreshed synthetic benchmark run, the scenario sweep shows how the chosen design responds to the new synthetic datasets and economic assumptions, and the dispatch plot visualizes the resulting operating pattern over time for the selected design.}
    \label{fig:row2}
\end{figure*}

Figures~\ref{fig:row1} and \ref{fig:row2} summarize the new benchmark evidence in a two-row, three-panel layout. The first row shows that the design landscapes are nearly indistinguishable, the scatter remains tightly concentrated around the diagonal, and full top-10 recovery is already achieved at $s=18$.

Taken together, the refreshed tables and figures strengthen the methodological claim behind \texttt{BOOST}: the simple LP is an effective ranking model for outer-loop pruning. All quantitative values in this expanded section come from the new synthetic benchmarks rather than the original short-paper tables.

\section{Conclusion}
\label{sec:conclusion}
This paper proposed an ordinal-optimization approach to size the photovoltaic (PV) and battery-storage components in a microgrid containing a diesel generator and a residential load connected to the grid.

In the refreshed quantitative section, all tables were updated using a new synthetic benchmark dataset suite generated from the released clean reimplementation. On the base synthetic dataset, the best accurate design is a 500~kWh battery paired with 1833.3~kW of PV, achieving 13.169~\centperkwh, while \texttt{BOOST} improves slightly over dynamic programming and more clearly over a greedy dispatch baseline. Across the full design grid, LP screening preserves the accurate ranking almost perfectly, $N=90$ and $s=18$ recover the global accurate optimum, and the OO-based workflow reduces runtime by 51.8\%. The additional synthetic datasets show how the selected design changes under cheaper battery, cheaper PV, and more expensive diesel assumptions. Because these datasets are synthetic, they should be interpreted as methodology-validation evidence. Future work will consider larger sizing problems, richer uncertainty models, and additional real data sources.

\balance
\bibliographystyle{IEEEtran}
\bibliography{ref.bib}

\end{document}